\documentclass[11pt,twoside]{article}
\usepackage{newpasp}
\usepackage{epsf}
\def\edcomment#1{\iffalse\marginpar{\raggedright\sl#1\/}\else\relax\fi}
\reversemarginpar

\begin{document}
\title{Are Cluster Radio Relics Revived Fossil Radio Cocoons?}
 \author{Torsten A.  En{\ss}lin} \affil{Max-Planck-Institut f\"{u}r
 Astrophysik, Karl-Schwarzschild-Str.1, 85740 Garching, Germany}
 \author{Gopal-Krishna} \affil{National Centre for Radio Astrophysics,
 Tata Institute of Fundamental Research, Pune University Campus,
 Ganeshkhind, Pune 411007, India}

\begin{abstract}
A new model for the, so called, {\it cluster radio relics} is
presented (En{\ss}lin \& Gopal-Krishna 2000). Fossil radio cocoons,
resulting from the former activity of radio galaxies, should contain a
low energy relativistic electron population and magnetic fields. Electrons with an age of even up
to 2 Gyr can be re-accelerated adiabatically to radio emitting
energies, if the fossil radio plasma gets compressed in an
environmental shock wave. Such a wave can be caused by a merging event of
galaxy clusters, or by the accretion onto galaxy clusters.  An
implication of this model is the existence of a population of diffuse,
ultra-steep spectrum, very low-frequency radio sources located inside
and possibly outside of clusters of galaxies, tracing the revival of
aged fossil radio plasma by the shock waves associated with
large-scale structure formation.
\end{abstract}
\setlength{\partopsep}{0em}
\setlength{\parskip}{0em}
\setlength{\parsep}{0em}
\vspace{-1em}
\section{Introduction}
\vspace{-0.5em}
The radio cocoons of radio galaxies become rapidly undetectable after
the central engine of the active galactic nucleus ceases to inject
fresh radio plasma.  Although undetectable, the aged radio plasma
should still be an important component of the inter-galactic medium
(IGM). We investigate the possibility of reviving patches of such
fossil radio plasma (also called {\it radio ghosts}, En{\ss}lin
(1999)) by compression in a shock wave produced in the IGM by the
flows associated with cosmological large-scale structure
formation. Such shock waves re-energize the electron population in the
fossil radio cocoons, which can lead to observable synchrotron
emission. This is probably the mechanism responsible for the, so
called, {\it cluster radio relics}, which are patches of diffuse,
sometimes polarized radio emission typically found at peripheral
locations in clusters of galaxies. These cluster radio relics can not
be simply relic radio galaxies, as their name suggests.  The spectral
ages of the electron population are {usually} too short to admit even
the nearest galaxy to have been the parent radio galaxy, which has
moved to its present location with a velocity typical for cluster
galaxies. Therefore, a recent enhancement of the nonthermal radio
output of the cluster relic sources is mandatory.


The connection between the presence of a shock wave and the appearance
of the cluster radio relic phenomenon was assumed to be due to Fermi-I
shock acceleration of electrons, in En{\ss}lin et al. (1998) and
Roettiger et al. (1999). While this process might work within the
normal IGM, several arguments favor the possibility that indeed old
fossil radio plasma is revived in the case of cluster relic sources:
\begin{itemize}
\setlength{\topsep}{0em}
\setlength{\partopsep}{0em}
\setlength{\parskip}{0em}
\setlength{\parsep}{0em}
\setlength{\itemsep}{0em}
\setlength{\itemindent}{0em}
\setlength{\leftmargin}{0em}
\setlength{\leftmargin}{0em}
\item
Cluster radio relics are extremely rare, whereas shock waves should be
very common within clusters of galaxies. The {dual} requirement of
a shock wave and fossil radio plasma, {for producing} a
cluster radio relic, {would} be an attractive explanation for the rareness of 
the relics.
\item
Fossil radio plasma with existing relativistic electron population and
fairly strong magnetic field appears to have ideal properties for being
brightened up during the shock's passage.
\item
The radio relic 1253+375 near the Coma cluster of galaxies
appears to be fed with radio plasma by the nearby galaxy NGC 4789
(En{\ss}lin et al. 1998).
\end{itemize}
But, if indeed the fossil radio plasma and not the normal IGM were to
become radio luminous at a shock wave, the expected very high
sound velocity of that relativistic plasma should forbid the shock in
the ambient medium from penetrating into the radio plasma. Thus, shock
acceleration is not expected to occur there. Instead, the fossil radio
plasma would get adiabatically compressed, and the energy gain of the
electrons is expected to be mainly due to adiabatic heating. It is the
purpose of this work to demonstrate that this process is sufficient to
account for the cluster radio relics.
\vspace{-1em}
\section{The Model\label{sec:phases}}
\vspace{-0.5em}
Between the release from a radio galaxy and the reappearance as a
cluster radio relic the radio plasmon undergoes several different
phases of expansion and contraction:\\
%
{\bf Phase 0: Injection.} The radio galaxy is active and a large
expanding volume is being filled with radio plasma. The expansion of
this cocoon is likely to be supersonic with respect to the outer
medium\\
{\bf Phase 1: Expansion.} After the central engine of the radio galaxy
becomes inactive, the radio cocoon might still be strongly
over-pressured compared to its gaseous environment and expands until
pressure equilibrium with environment is reached.  The radio cocoon
probably becomes undetectable during this phase, and therefore becomes
a fossil radio cocoon or a so called radio ghost.\\
{\bf Phase 2: Lurking.} Due to the previous adiabatic energy losses of the
electrons, they are at low energies. Their radiation losses, which
strongly depend on the particle energies, are therefore strongly
diminished. Furthermore, the synchrotron losses are
reduced due to the weaker magnetic field during the expanded state of
the radio cocoon. The adiabatic losses are reversible, and will be
reversed during the subsequent compression phase.\\
{\bf Phase 3: Flashing.} The fossil radio plasma gets dragged into a
environmental shock wave. The relativistic electron population and the
magnetic field gain energy adiabatically, leading to a steep
enhancement of the synchrotron emissivity. The fossil appears as a
`relic'.\\
{\bf Phase 4: Fading.} The radio plasma is in pressure equilibrium
with the post-shock ambient medium. The radio emission of the relic 
now fades away due to the heavy radiation losses.

We followed the evolution of a cocoon of radio plasma for different
scenarios (A, B, and C) and calculated the resulting radio emission.
In A, the relic is located at the cluster center, while in
B, its location is near the cluster boundary, i.e., in the proximity
of the accretion shock wave. In both scenarios the duration of phase 2
was chosen to be so long that the shocked radio plasma could be barely
observed as a weak ultra-steep spectrum source: 0.1 Gyr in A and 1 Gyr
in B. Scenario C demonstrates that a shorter phase 2 can lead to a
moderately steepened spectrum of the cluster radio relic.  The
different scenarios and the calculations of the radio spectra are
described in detail in En{\ss}lin \& Gopal-Krishna (2000). The
resulting radio spectra of the radio fossil/relic in scenario A are
displayed in Fig. 1 together with the spectrum of an observed relic.

\begin{figure}[t]
\begin{center}
\plottwo{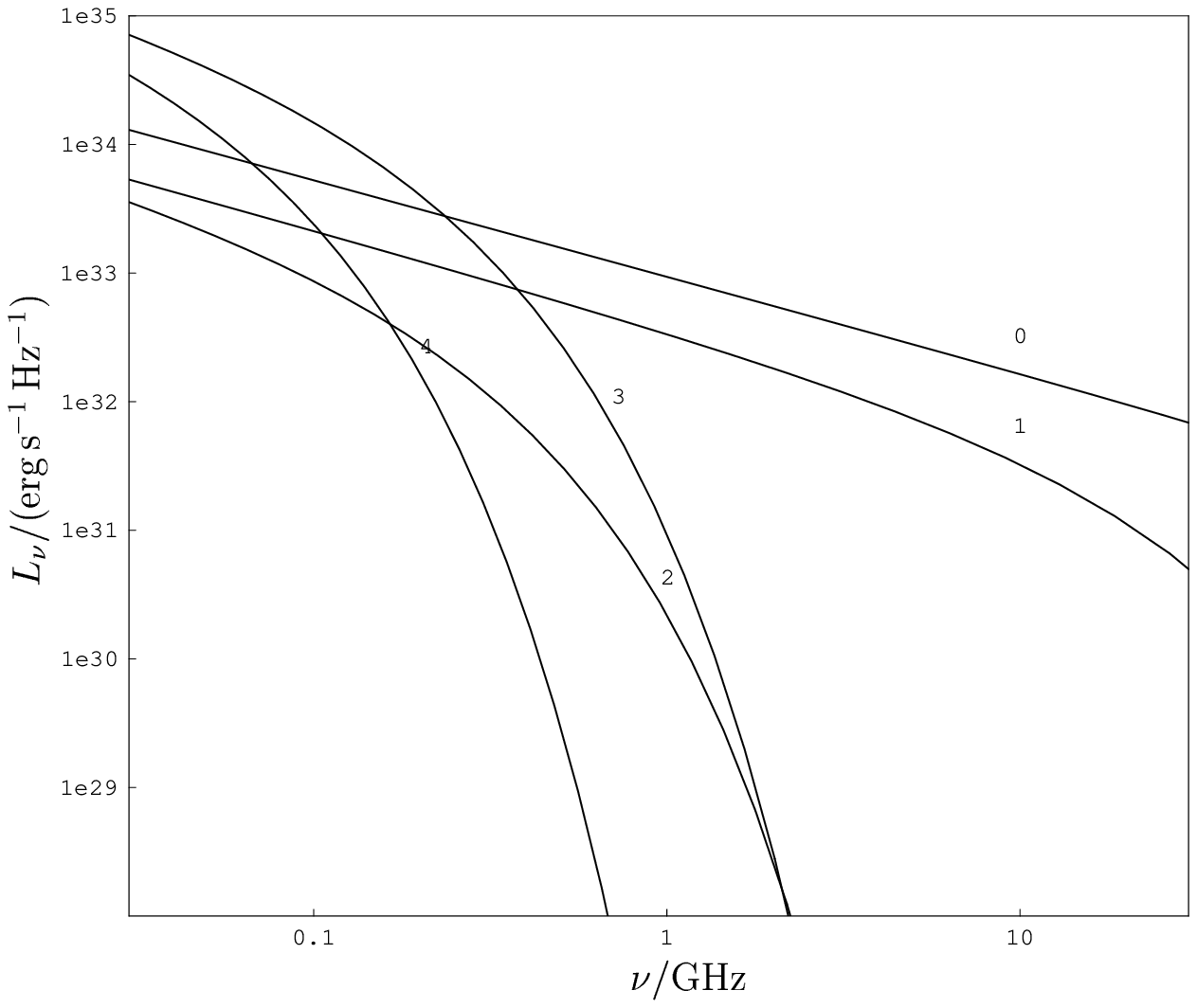}{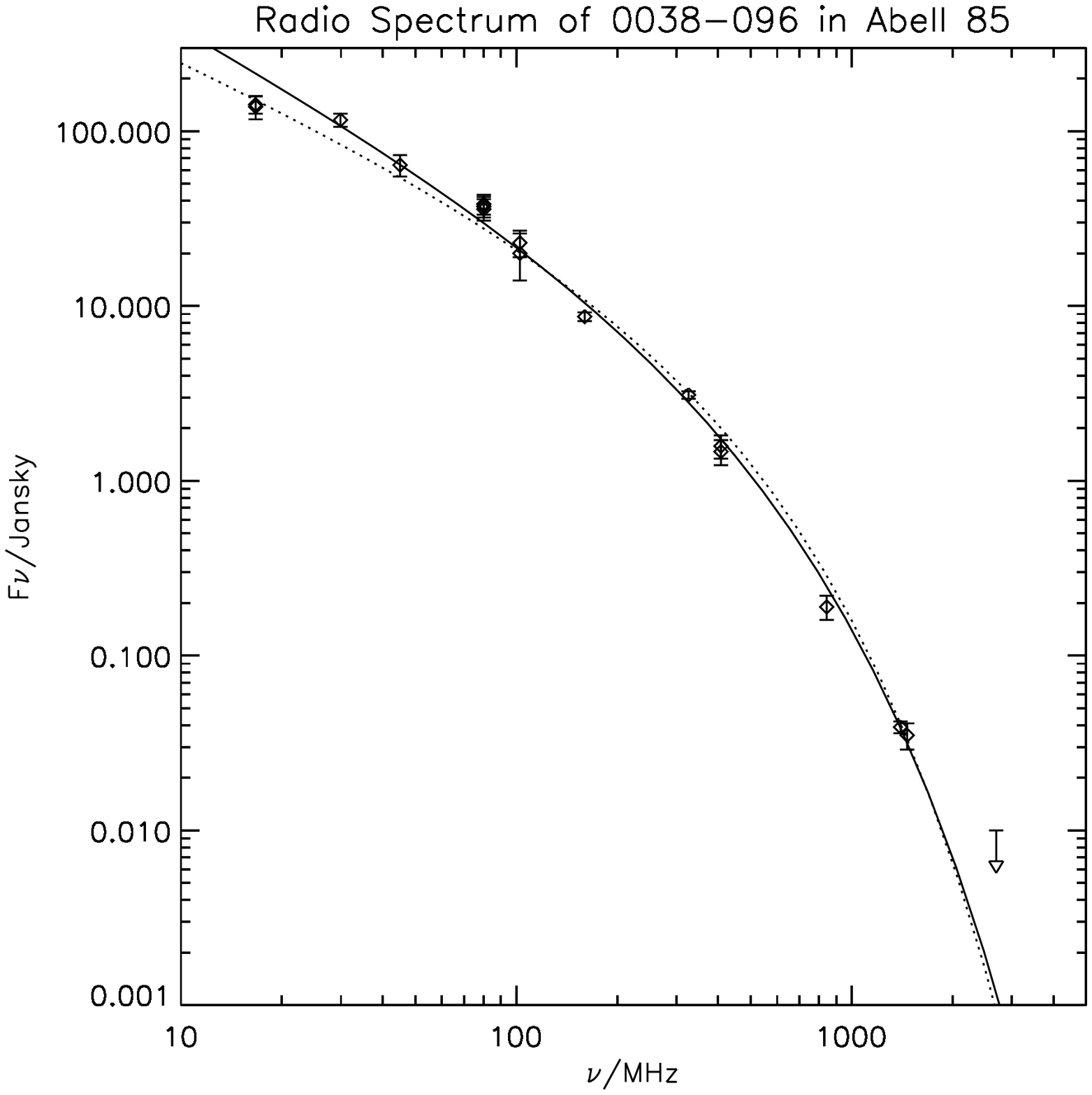}
\end{center}
\caption[]{\label{fig:syncA} Left: Radio spectrum of the radio cocoon
in scenario A at the end of phases 0-4. Right: \label{fig:A85} Radio
spectrum of the radio relic in A85, compared to spectra resulting from
this model (lines).}
\vspace{-1em}
\end{figure}
\vspace{-1em}
\section{Discussion\label{sec:discussion}}
\vspace{-0.5em}
We have argued here that adiabatic compression in cluster shocks can
revive fossil radio plasma to radio detection, even up to 2 Gyr after
the cessation of the nuclear activity 
and thus explain the observed
cluster radio relics.  Below, we summarize some merits of this
model (also, see Sect 1.):
\begin{itemize}
\setlength{\topsep}{0em}
\setlength{\partopsep}{0em}
\setlength{\parskip}{0em}
\setlength{\parsep}{0em}
\setlength{\itemsep}{0em}
\setlength{\itemindent}{0em}
\setlength{\leftmargin}{0em}
\setlength{\leftmargin}{0em}
\item The observed connection of cluster radio relics to shock waves
arises naturally in this model.
\item The presence of the radio galaxy NGC 4789 and the morphological
connection of its radio tails to the relic 1253+275 seems to be the
{\it smoking gun} of cluster radio relic formation by compression of
fossil radio plasma.
\item The expected very high sound speed within the radio plasma should
virtually forbid shock waves of the ambient medium to penetrate into
radio cocoons. This renders adiabatic compression as a more plausible
means of reviving the fossil plasma.
\item Cluster radio relics are rare. Not only a shock wave and fossil
radio plasma are both required in this model, in order to produce a
cluster radio relic, but also the fossil radio plasma cannot grow too
old (of the order of 0.1 Gyr in the center of clusters and 1 Gyr at
their boundaries). Otherwise, the depletion of the high energy end of
the electron population would be too severe for the adiabatic gains
during the compression phase to be able to shift them to radio
emitting energies. Since radio fossils should be very common
(En{\ss}lin \& Kaiser 2000), these requirements help to explain the
observed rarity of cluster radio relics.
\item The observed tendency of relics to appear at peripheral
locations in clusters follows in this model from the much shorter
synchrotron lifetimes of the radio emission and, consequently, much
shorter ages which the revive-able fossil radio cocoons can have near
the centers.
\end{itemize}
We conclude that the model presented here provides a fairly natural
and promising explanation for the phenomenon of cluster radio
relics. It predicts the existence of a population of diffuse,
ultra-steep, very low frequency radio sources inside and possibly also
outside of clusters of galaxies, due to the age dependence of the
upper frequency cutoff of the radio emission arising from the revivable
fossil radio plasma. The ongoing rapid improvements in the
sensitivities at low radio frequencies (Giant Meterwave Radio
Telescope: (GMRT), Low Frequency Array (LOFAR)) may therefore open a
new window on cosmological structure formation by detecting shock
waves marked by such relic radio sources.


\def\nat{Nature}
\def\aj{AJ}
\bibliography{tae}
\bibliographystyle{aabib99}



\end{document}